\documentclass[aps, prd, twocolumn, showpacs, floatfix, letterpaper,
  nofootinbib, superscriptaddress,longbibliography]{revtex4-1}

\usepackage{graphicx}
\usepackage{epsfig}
\usepackage{bm}
\usepackage{amsfonts}
\usepackage{pgf}
\usepackage{color}
\usepackage[T1]{fontenc}
\usepackage[latin1]{inputenc}
\usepackage{graphicx}
\usepackage[english]{babel}
\usepackage{amsmath}
\usepackage{amssymb}
\usepackage{amsfonts}
\usepackage{makecell}
\usepackage{hyperref}
\usepackage[draft,deletedmarkup=xout]{changes}
\usepackage{mathtools}
\usepackage{soul}
\usepackage{ulem}
\usepackage{relsize}
\usepackage{enumitem} 

\definecolor{green}{rgb}{0.19,0.64,0.54}
\definecolor{blue}{rgb}{0,0,1} \definecolor{reddish}{rgb}{0.65, 0.2,
  0.2} \definecolor{darkgreen}{rgb}{0.2,0.7,0.3}
\definecolor{darkblue}{rgb}{0.3,0.40,0.48}
\definecolor{gray}{rgb}{.8,.8,.8}

\hypersetup{,
  pdftitle={Time},
  pdfauthor={A Boldrin, P Malkiewicz and P Peter},
  colorlinks=true, % false: boxed links; true: colored links
  linkcolor=darkblue, % color of internal links
  citecolor=darkgreen, % color of links to bibliography
  filecolor=reddish, % color of file links urlcolor=blue, % color of external links
  linktocpage=true
  }

\newcommand{\dd}{\mathrm{d}}
\newcommand{\ex}{\mathrm{e}}

\newcommand{\ud}{\mathrm{d}}

\newcommand{\be}{\begin{equation}}
\newcommand{\ee}{\end{equation}}
\newcommand{\ba}{\begin{eqnarray}}
\newcommand{\ea}{\end{eqnarray}}

\begin{document}

\title{Time problem in primordial perturbations}

\author{Alice Boldrin} \email{Alice.Boldrin@ncbj.gov.pl}

\affiliation{National Centre for Nuclear Research, Pasteura 7, 02-093
  Warszawa, Poland}

\author{Przemys{\l}aw Ma{\l}kiewicz}
\email{Przemyslaw.Malkiewicz@ncbj.gov.pl}

\affiliation{National Centre for Nuclear Research, Pasteura 7, 02-093
  Warszawa, Poland}

\author{Patrick Peter} \email{patrick.peter@iap.fr}

\affiliation{${\cal G}\mathbb{R}\varepsilon\mathbb{C}{\cal
      O}$ -- Institut d'Astrophysique de Paris, CNRS and Sorbonne
  Universit\'e, UMR 7095 98 bis boulevard Arago, 75014 Paris, France}

\begin{abstract}
We study the nonunitary relation between quantum gravitational models
defined using different internal times. We show that, despite the
nonunitarity, it is possible to provide a prescription for making
unambiguous, though restricted, physical predictions independent of
specific clocks. To illustrate this result, we employ a model of
quantum gravitational waves in a quantum Friedmann universe.
\end{abstract}

\maketitle

\section{Introduction}

Quantum gravity models suffer from the infamous time problem
\cite{Kuchar:1991qf, Isham:1992ms, Anderson:2010xm,Kiefer:2021zdq} as
the external and absolute time on which nonrelativistic physics is
based, is absent in Einstein's theory of gravity. Therefore, one has
to rely on largely arbitrary physical variables, known as internal
time variables or internal clocks, to follow changes occurring in
gravitational systems. By virtue of the principle of general
relativity (time-reparametrization invariance), the free choice of
internal time variable has no physical consequence in the classical
theory. Upon passing to quantum theory, however, different choices of
internal time variables are known to produce unitarily inequivalent
quantum models \cite{Deser:1961zza, Beluardi:1994jb, Hajicek:1999ti,
  Catren:2000br, Hajicek:2000dy,Malkiewicz:2014fja,
  Malkiewicz:2017cuw,Bojowald_2018}. The problem of finding the
correct interpretation of these nonequivalent models is commonly
referred to as the time problem.

In this article, we look for the most plausible interpretation of such
nonequivalent clocks. Our analysis is based on the model of primordial
gravitational waves propagating across the Friedmann universe. It is
important to note that similar models were previously used for making
predictions for the primordial amplitude spectrum of density
perturbation, which are greatly constrained by observations (see,
e.g., Refs.~\cite{Peter:2006hx, Pinto-Neto:2008gdo,
  Peter:2008qz}). Remarkably, to the best of our knowledge, the time
problem has never been studied for such models, so it is important to
clarify the role and the interpretation of internal time variables in
their dynamics. We expect that the ensuing conclusions should equally
apply to all cosmological models.

The fact that the dynamics are unitarily inequivalent in different
clocks is widely known and well documented with plenty of examples;
see, e.g., Refs.~\cite{Malkiewicz:2016hjr, Malkiewicz:2019azw,
  Malkiewicz:2022szx}. In this context, it is sometimes emphasized
that the only measurable quantities in quantum gravity are
gauge-invariant variables that do not depend on the employed clock
\cite{Rovelli:2001bz, Dittrich:2004cb}. They are constants of
motion. However, they are said to encode all the relational dynamics
in spite of being nondynamical themselves. From this point of view,
dynamical quantities are not fundamental and are ambiguously given by
one-parameter families of gauge-invariant quantities, with each family
representing the motion with respect to a specific internal time. The
differences are seen as natural rather than inconsistencies that
should be worried about. From our viewpoint, on the other hand, the
dynamical variables can serve as fundamental variables, and the
differences in their dynamics call for a careful interpretation,
before allowing for physical predictions.

The cosmological system examined in this article exhibits, as
expected, dynamical discrepancies when based on different clocks. The
discrepancies concern both the background and perturbation
variables. This leads us to ask a fundamental question: what are the
dynamical predictions of quantum cosmological models, that do not
depend on the employed time variable?

We address the above question within the reduced phase space
quantization. Namely, we solve the Hamiltonian constraint and choose
the internal time variable prior to quantization. An alternative
approach would be to first quantize and then solve the constraint
quantum mechanically while promoting one of the variables as internal
time. Both approaches lead to the same time problem and, therefore,
using the technically less involved reduced phase approach is well
justified (see, however, Ref.~\cite{Bojowald_2023} for recent
developments in the alternative approach). Most significantly, within
the reduced phase space approach, there exists a theory of clock
transformations, which is completely crucial for the purpose of this
work \cite{Malkiewicz:2015fqa}. Thanks to these precisely defined
transformations, we are able to explore all possible clocks and
quantize them with an assumption of fixed operator ordering. Hence,
any quantum ambiguities found arise from the differences between
clocks rather than the differences between quantization prescriptions.

The outline of this article is as follows. In Sec. \ref{sec2}, we make
a brief introduction to the theory of clock transformations in the
reduced phase space of gravitational models. We explain how this
theory allows one to remove irrelevant quantization ambiguities when
passing to quantum theory based on different clocks and with different
basic dynamical variables. In Sec. \ref{sec3}, we formulate the
reduced phase space description of the Friedmann universe with
gravitational waves with respect to a fluid time and obtain the
general clock transformation. In Sec. \ref{sec4}, we quantize our
model and establish a convenient semiclassical approximation. Section
\ref{sec5} deals with concrete clock transformations applied to our
model and makes comparisons between the resulting dynamics. We
summarize our findings, discuss their plausible interpretation, and
suggest some directions to move forward in Sec. \ref{sec6}.

\section{Clock transformations in totally constrained systems}
\label{sec2}

One crucial characteristic feature of canonical relativity is the
appearance of the Hamiltonian constraint; it is a consequence of the
fact that the dynamics of three surfaces is generated by infinitesimal
timelike diffeomorphisms, and the latter leave the full
four-dimensional spacetime invariant. It by no means makes the
dynamics of three surfaces spurious or redundant. Indeed, the
Hamiltonian constraint dynamics is a feature of any canonical
relativistic theory of gravity, be it Einstein's or any modified
gravity theory, though their dynamics are different. The correct
interpretation of canonical relativity assumes the lack of an
absolute, external time in which three surfaces evolve, and replaces
it with internal variables that serve as clocks in which the dynamics
of three surfaces takes place. None of the internal clocks can play a
privileged role as the principle of relativity states. This picture is
certainly self-consistent in the classical theory. At the quantum
level, no spacetime exists and, as we will see later, the principle of
relativity takes a somewhat altered form. In order to study it, we
need to extend the canonical formalism by including clock
transformations that transform a canonical description from one
internal clock to another; only then can we move to the quantum level
where these new transformations become a key to unlock the principle
of quantum relativity.

Let us consider a system consisting of a set of $N+1$ canonical
variables $\{q_\alpha,p^\alpha\}_{\alpha= 0,\cdots, N}$ and assume a
Hamiltonian constraint taking the form
$$C(q_\alpha,p^\alpha)\approx 0,$$ where ``$\approx$'' is the weak
equality in the Dirac sense~\cite{dirac1964lectures}. Suppose that one
of the positions, say $q_0$, varies monotonically with the evolution
generated by the constraint, i.e., $\forall \, q_0, \{q_0,C\}_\textsc{pb}
\neq 0$. It is then possible to assign to $q_0$ the role of an
internal clock in which the evolution of the remaining variables
occurs. This evolution is then governed by a Hamiltonian that is not a
constraint. At this stage, it may seem that the time variable is fixed
once and for all, which would contradict the principle of relativity;
we discuss below in what sense this is not the case.

The reduced Hamiltonian formalism is obtained from the initial
symplectic form $\Sigma = \dd q_\alpha \wedge \dd p^\alpha$ (Einstein
convention assumed), evaluated on the constraining surface, namely,
\begin{equation}
\begin{gathered}
\Sigma\big|_{C=0} = \left(\ud q_I\wedge \ud p^I+\ud q_0\wedge \ud
p^0\right)\big|_{C=0}\\
\null =\ud q_I \wedge \ud p^I-\ud t\wedge \ud H,
\label{OmegaC}
\end{gathered}
\end{equation}
where $I=1,\cdots, N$, and $H=H\left( q_0,q_I,p^I\right)$ is the
nonvanishing reduced Hamiltonian such that $p_0+H\approx 0$. Note that
both $q_0$ (denoted by $t$ from now on to emphasize its role as a time
variable) and $p^0$ are removed from the phase space and the remaining
dynamical variables are no longer constrained. Indeed, their dynamics
reads
$$\frac{\ud q_I}{\ud t}=\frac{\partial H}{\partial p^I}\qquad
\hbox{and} \qquad \frac{\ud p^I}{\ud t}=-\frac{\partial H}{\partial
  q_I},$$
which is entirely solved once an arbitrary initial condition
$(q_I^\text{ini},p^I_\text{ini}, q_0^\text{ini})$ is provided.

In order to restore the principle of relativity, we need to allow for
any clock, denoted by $\tilde{t}$, which monotonically varies with the
evolution generated by the constraint $\{\tilde{t},C\}_\textsc{pb}
\neq 0$. This new clock must be a function of the old clock and the
old canonical variables, $\tilde{t}=\tilde{t}(q_I,p^I, t)$. Thus, it
must satisfy
\begin{equation}
\frac{\dd \tilde{t}}{\dd t} = \frac{\partial \tilde{t}}{\partial t} +
\underbrace{\frac{\partial \tilde{t}}{\partial q_I} \frac{\partial
    H}{\partial p^I} - \frac{\partial \tilde{t}}{\partial p^I}
  \frac{\partial H}{\partial q_I}} _{\{\tilde{t},H \}_\textsc{pb}}
\neq 0.
\end{equation}
The original symplectic form induced on the constraint surface $C=0$
must read in some new canonical variables:
$$
\Sigma\big|_{C=0}=\ud \tilde{q}_I\wedge\ud \tilde{p}^I - \ud\tilde{t}
\wedge\ud \tilde{H},
$$
so that the new reduced formalism is still canonical. This implies
that there must exist an invertible map between the old and the new
variables:
\begin{equation}
\tilde{t}=\tilde{t}(q_I,p^I, t),\ 
\tilde{q}_I=\tilde{q}_I(q_J,p^J,t), \
\tilde{p}^I=\tilde{p}^I(q_J,p^J, t),
\label{trans}
\end{equation}
and the natural question to ask is whether these transformations are
canonical. In principle, and in all the relevant cases, they most
certainly are not. It can be shown that clock transformations form a
group of generally noncanonical transformations with canonical
transformations as its normal subgroup \cite{Malkiewicz:2016hjr};
finding them is, in general, a difficult task. However, for an
integrable dynamical system, the problem can be reduced to that of
solving a set of algebraic equations.

If a dynamical system is integrable, then we may find a complete set
of canonical constants of motion, denoted by $D_I$. Let them be
functions of the old internal time and old canonical variables,
$D_I=D_I(q_J,p^J, t)$. Note that substituting back $t\rightarrow q_0$,
they must commute with the original constraint,
$\{D_I,C\left(q_\alpha,p^\alpha\right)\}_\textsc{pb} = 0$.  They are
therefore genuine Dirac observables in the constrained system. The new
internal time $\tilde{t}=\tilde{t}(q_I,p^I, t)$ and new canonical
variables can then be found according to the algebraic relations
\begin{align}
\tilde{t}=\tilde{t}(q_I,p^I, t),~~D_I(q_J,p^J,
t)=D_I(\tilde{q}_J,\tilde{p}^J, \tilde{t}),\label{TF}
\end{align} 
where we formally substitute the canonical variables in the
expressions for Dirac observables $D_I$, i.e., we assume the same
functional dependence of $D_I$ in both sets of variables. The number
of $D_I$ is equal to the number of the new canonical variables
$\tilde{q}_J$ and $\tilde{p}^J$, and thus, leaving aside singular
cases, the above relations determine $\tilde{q}_J$ and $\tilde{p}^J$
completely. The result is a new canonical formalism based on a new
internal clock. Let us note that, by virtue of Eq.~\eqref{TF}, if a
solution to the dynamics is known in one clock, i.e., $t\rightarrow
\left[ q_I\left(D_J,t\right),p^I\left(D_J,t\right) \right]$, then it
is readily known for all other clocks and reads $\tilde{t}\rightarrow
\left[\tilde{q}_I=q_I\left(D_J,\tilde{t}\right),
  \tilde{p}^I=p^I\left(D_J,\tilde{t}\right)\right]$. This makes the
choice of the new canonical variables $\tilde{q}_I$ and $\tilde{p}^I$
via Eq.~\eqref{TF} very convenient: the formal description of the
system is the same in all clocks, only the physical meaning of the
clock and basic variables changes, which is emphasized by the use of a
tilde (~$\tilde{}$~) over the variable names.

The use of Dirac observables in the derivation of clock
transformations gives an invaluable advantage when passing to quantum
theory. Our goal is to make a comparison between quantum theories
based on different internal clocks of a single physical
system. Therefore, it is of uttermost importance to make sure that the
theories are different only insofar as their clocks differ, and not
due to other quantization ambiguities such as the well-known factor
ordering. This state of affairs can be achieved by fixing a quantum
representation of the Dirac observables and then defining basic and
compound observables as functions of the quantum Dirac observables,
both in the original
\begin{equation*}
\widehat{q}_I=q_I(\widehat{D}_J,t), \qquad
\widehat{p}^I=p^I(\widehat{D}_J,t),
\end{equation*}
and the new variables
\begin{equation*}
\widehat{\tilde{q}}_I={q}_I(\widehat{D}_J,\tilde{t}),
\qquad \widehat{\tilde{p}}^I = {p}^I(\widehat{D}_J,\tilde{t}).
\end{equation*}
These definitions imply that ${q}_I$ and ${p}^I$ are promoted to the
same operators as ${\tilde{q}}_I$ and ${\tilde{p}}^I$,
respectively. We invert this reasoning and start by assuming the same
operators for ${q}_I$ and ${\tilde{q}}_I$ as well as ${p}^I$ and
${\tilde{p}}^I$. This implies that the Dirac observables being the
same functions in both sets of basic variables are promoted to the
same operators irrespective of the choice of clock. Hence, the quantum
descriptions in different clocks are formally the same; only the
physical meaning of the basic operators changes from one clock to
another, which is emphasized by the use of tilde.  Obviously, a unique
ordering prescription has to be used in all the above formulas. In
principle, after this step, any physically interesting aspect of the
quantum theories can be compared. In the following section, we
introduce the model on which we discuss such comparisons.

\section{Canonical cosmological model}
\label{sec3}

We consider a flat Friedmann-Lema\^{\i}tre-Robertson-Walker (FLRW)
universe filled with radiation and perturbed by gravitational waves;
the line element of the model reads (in units such that $c=1$)
$$\ud s^2=-N^2(t)\ud t^2+a^2(t)\left[ \delta_{ij}+
h_{ij}(\bm{x},t)\right]
\ud x^i\ud x^j,$$ where $h_{ij}$ represent the
gravitational waves (tensor perturbations); it satisfies
$h_{ij}\delta^{ij}=0$ and $\partial^j h_{ij}=0$.
Finally, we assume a toroidal spatial topology
with each comoving coordinate $x^i\in [0,1)$. Setting
$N\to a$ means one considers the conformal time; we shall
henceforth denote it by $\eta$ to agree with most of the
cosmology literature.

\subsection{Perturbative Hamiltonian}

Let us now build the canonical description of these gravitational
waves in an FLRW universe. The relevant canonical variables are the
scale factor $a$ and its conjugate momentum $p_a$ to describe the
background, while the tensor perturbations are represented by the
gravitational wave amplitude $\mu^{(\lambda)} = a h^{(\lambda)}$ and
its conjugate momentum $\pi^{(\lambda)}$, with $\lambda \in \{ +,
\times\}$ and $h_{ij} = \sum_\lambda h^{(\lambda)}
\varepsilon_{ij(\lambda)}$ (see, e.g.,
Refs.~\cite{Peter:2013avv,Micheli:2022tld} for details on the helicity
expansion).

The matter component is assumed to be a radiation fluid with energy
density $p_0$ conjugate to a timelike variable $q_0$. The
gravitational constraint is expanded to second order through
$$
H_\text{tot} = H^\text{(b)} + \sum_{\bm{k}}
H^\text{(p)}_{\bm{k}}
$$ (recall the spatial sections are compact), with the background
Hamiltonian given by
\begin{equation}
H^\text{(b)} = -\frac12 p_a^2-p_0.
\label{H0}
\end{equation}
At this stage, one can identify the internal time $q_0$ with the
conformal time $\eta$ as it reduces the zeroth-order Hamiltonian into
\begin{equation}
\begin{gathered}
\Sigma\big|_{H^{(0)}=0}
=\left(\ud a\wedge\ud p_a+\ud q_0\wedge \ud
p_0\right)\big|_{H^\text{(b)}=0}
\\
=\ud a\wedge\ud p_a-\ud \eta\wedge\ud
\left( \frac{1}{2}p_a^2 \right),
\label{OmegaCbckg}
\end{gathered}
\end{equation}
leading to the physical zeroth-order Hamiltonian 
\begin{equation}
H^{(0)} = \frac12 p_a^2,
\label{H0phys}
\end{equation}
while preserving the form of the perturbation Hamiltonian
$H^\text{(p)}_{\bm{k}}$. The latter reads, at second order
\begin{equation}
H^\text{(p)}_{\bm{k}} \to H^{(2)}_{\bm{k}} = 
- \sum_{\lambda=+,\times} H^{(2)}_{\bm{k},\lambda}
\label{Hpert}
\end{equation}
with
\begin{equation}
H^{(2)}_{\bm{k},\lambda} = \frac12 \left| \pi^{(\lambda)}_{\bm{k}}
\right|^2 + \frac12 \left( k^2 - \frac{a''}{a} \right) \left|
\mu^{(\lambda)}_{\bm{k}} \right|^2,
\label{H2kini}
\end{equation}
where a prime stands for a derivative with respect to the conformal
time. Since the tensor perturbations are real, one has $\mu^{(\lambda)
  *}_{\bm{k}} = \mu^{(\lambda)}_{-\bm{k}}$.  Moreover, since the
background is isotropic, one can restrict attention to upward directed
wave vectors $\bm{k}$ by merely canceling the factor $\frac12$ in
$H^{(2)}_{\bm{k},\lambda}$.  This permits one to write the final
second-order Hamiltonian as
\begin{equation}
H^{(2)}_{\bm{k},\lambda} = 
\pi^{(\lambda)}_{\bm{k}} \pi^{(\lambda)}_{-\bm{k}}
+ \left( k^2 - \frac{a''}{a} \right)
\mu^{(\lambda)}_{\bm{k}} \mu^{(\lambda)}_{-\bm{k}}.
\label{H2k}
\end{equation}
Note that, for the radiation fluid we are concerned with here, the
Hamiltonian \eqref{H0phys} yields as equations of motion $p_a = a'$
and $p_a'=0$, thus leading to $a''=0$: the potential for producing
gravitational waves is indeed classically vanishing if the universe is
radiation dominated.

Determining the solution to the dynamics of gravitational waves is
straightforward in the radiation case. While it is possible to
consider a general fluid with an arbitrary barotropic index $w$ (this
case can be solved analytically in terms of Bessel functions, see,
e.g., \cite{Mukhanov:1990me}), such a consideration is not relevant to
the objectives of this work. We expect that the clock effects obtained
below are not specific to any matter content but must be present
whenever quantum uncertainties in the background geometry are taken
into account. In fact, it can be argued that, since gravitational
waves are affected by the choice of the equation of state only insofar
as the background time development depends on it through
Eq.~\eqref{H2k}, our results should qualitatively hold, if not
quantitatively, for all physically relevant choices of $w$.

\subsection{Dirac observables}

Now we shall find the constants of motion that form canonical
pairs. To this end, we need to solve the partial differential
equations
\begin{equation}
\frac{\ud D}{\ud \eta}=\frac{\partial D}{\partial
\eta}+\{D,H^{(0)}+H^{(2)}\}_\textsc{pb}=0.
\label{Diracs}
\end{equation}
At zeroth order, this is
$$
\frac{\partial D}{\partial \eta} + p_a
\frac{\partial D}{\partial a} = 0,
$$
with solutions
\begin{equation}
D_1=a - p_a \eta \qquad \hbox{and} \qquad
D_2 = p_a.
\label{D1D2}
\end{equation}
At first order, Eq.~\eqref{Diracs} reads
$$ \frac{\partial \delta D}{\partial \eta} + p_a \frac{\partial \delta
  D}{\partial a} = \pi^{(\lambda)}_{\bm{k}} \frac{\partial \delta D}
{\partial \mu^{(\lambda)}_{\bm{k}}} -k^2 \mu^{(\lambda)}_{\bm{k}}
\frac{\partial \delta D}{\partial \pi^{(\lambda)}_{\bm{k}}},
$$
where we considered the classical solution $a''=0$. Since we are
considering only first-order perturbations, we demand that $\delta D$
be linear in the perturbation variables $\mu^{(\lambda)}_{\bm{k}}$ and
$\pi^{(\lambda)}_{\bm{k}}$. The lhs of the above equation is greatly
simplified if $\delta D$ depends only on the variable $y=\eta +
a/p_a$, so we look for a solution of the form $\delta D^{(\lambda)} =
\mu^{(\lambda)}_{\bm{k}} \alpha (y) + \pi^{(\lambda)}_{\bm{k}}
\beta(y)$, leading to
$$
2\frac{\dd \alpha}{\dd y} \mu^{(\lambda)}_{\bm{k}} +
2\frac{\dd \beta}{\dd y} \pi^{(\lambda)}_{\bm{k}}
= \alpha \pi^{(\lambda)}_{\bm{k}} 
- k^2 \beta \mu^{(\lambda)}_{\bm{k}}.
$$
Assuming independent variations of $\mu^{(\lambda)}_{\bm{k}}$ and
$\pi^{(\lambda)}_{\bm{k}}$, one gets $2\dd\alpha/\dd y=-k^2\beta$ and
$2\dd\beta/\dd y=\alpha$, and finally $4\dd^2\alpha/\dd
  y^2 = - k^2 \alpha$, so that, setting $$\Omega_k = \frac{k}{2}
\left( \eta + \frac{a}{p_a} \right),$$ one gets two independent
solutions for each polarization, or, in other words, four first-order
constants, reading
\begin{equation}
\begin{gathered}
\delta D^{(\lambda)}_{1,\bm{k}} = \sqrt{k}
\sin\Omega_k \, \mu^{(\lambda)}_{\bm{k}}
-\frac{\cos\Omega_k}{\sqrt{k}}\,
\pi^{(\lambda)}_{\bm{k}},\\
\delta D^{(\lambda)}_{2,\bm{k}} = \sqrt{k} \cos\Omega_k\,
\mu^{(\lambda)}_{\bm{k}} + \frac{\sin\Omega_k}{\sqrt{k}} \,
\pi^{(\lambda)}_{\bm{k}}.
\label{deltaD}
\end{gathered}
\end{equation}
In Eq.~\eqref{deltaD}, the normalization has been chosen so as to
ensure that all these Dirac observables indeed form canonical pairs,
namely
$$\left\{ D_1 , D_2 \right\}_\textsc{pb} = 1\qquad \hbox{and}\qquad
\left\{ \delta D^{(\lambda)}_{1,\bm{k}}, \delta
D^{(\bar{\lambda})}_{2,\bm{k}} \right\}_\textsc{pb} =
\delta_{\lambda\bar{\lambda}}.$$ From now on, we drop the index
$\lambda$ and consider just a single polarization mode
$(\mu_{\bm{k}},\pi_{\bm{k}})$.

\subsection{Clock transformations}\label{clocktra}

Having set the full model, and before moving on to its quantum
counterpart, let us first consider a general clock transformation
\begin{equation}
\eta \rightarrow \tilde{\eta} = \eta + \Delta(a,p_a,\eta),
\label{CT}
\end{equation}
where $\Delta$ is a delay function that, in general, varies between
the trajectories as well as along them.  At the background level,
implementing the recipe given by Eq.~\eqref{trans}, i.e.,
$D_{1,2}(a,p_a,\eta) = D_{1,2} (\tilde{a},\tilde{p}_a,\tilde{\eta})$,
to the transformation~\eqref{CT} yields
$$
a-p_a \eta = \tilde{a}-\tilde{p}_a \tilde{\eta} \quad
\hbox{and}\quad p_a = \tilde{p}_a,
$$
leading to
\begin{equation}
\tilde{a} = a + p_a \Delta  \quad
\hbox{and} \quad p_a = \tilde{p}_a.
\label{CTbckg}
\end{equation}
In order that the clock transformation \eqref{CT} actually defines a
new and physically acceptable clock, the delay function $\Delta$ must
be subject to two conditions. First, the new clock must run forward,
that is
\begin{equation}
\frac{\ud \tilde{\eta}}{\ud \eta} = 1 + 
\frac{\ud \Delta}{\ud \eta} = 1 + 
\frac{\partial \Delta}{\partial \eta} + 
p_a \frac{\partial \Delta}{\partial a}>0,
\label{Cond1}
\end{equation}
where in the second equality we used the zeroth-order Hamiltonian
$H^{(0)}$ given by Eq.~\eqref{H0phys} and the associated equations of
motion.

The second condition that a clock transformation must satisfy is that
the ranges of the basic variables $a$ and $p_a$ must be preserved,
thereby preventing the appearance of nontrivial ranges that may induce
new and potentially unsolvable quantization issues. This second
condition implies
\begin{subequations}\label{Cond2}
\begin{align}
    \lim_{p_a\rightarrow\pm\infty} \tilde{p}_a(a,p_a,\eta) &=
    \pm\infty,
    \label{Cond2a}
\\
\tilde{a}(a,p_a,\eta)\big|_{a=0}&=0.
\label{Cond2b}
\end{align}
\end{subequations}
The first equality \eqref{Cond2a} is trivially satisfied in the
present case because of \eqref{CTbckg}.  For $\Delta=\Delta(a,p_a)$,
to which we shall restrict attention in what follows, the second
equality \eqref{Cond2b} is identical to demanding that the delay
function at vanishing scale factor should also vanish,
$\Delta(0,p_a)=0$. This condition also ensures that the slow-gauge
clock is transformed into another slow-gauge clock, that is, the
boundary is reached within a finite amount of time (see
Ref.~\cite{Malkiewicz:2019azw}). Such a condition \eqref{Cond2b},
although irrelevant in the classical theory, is crucial for the
existence of a bounce at the quantum level where the clock must
smoothly connect contracting and expanding trajectories. Were
\eqref{Cond2} violated, the clock transformations would break the
bouncing trajectories.

It turns out that the condition \eqref{Cond1} is equivalent to the
existence of a one-to-one map between the reduced phase spaces
$(a,p_a)$ and $(\tilde{a},\tilde{p}_a)$, i.e., the determinant
\begin{equation}
\frac{\partial \left( \tilde{a},\tilde{p}_a\right)}
{\partial\left( a,p_a \right)} = \left|
\begin{array}{cc} \displaystyle\frac{\partial\tilde{a}}{\partial a} & 
\displaystyle\frac{\partial\tilde{a}}{\partial p_a} \\ &
\\ \displaystyle\frac{\partial\tilde{p}_a}{\partial a} &
\displaystyle\frac{\partial\tilde{p}_a}{\partial p_a}
\end{array}\right|> 0,
\label{Jac}
\end{equation}
which is indeed Eq.~\eqref{Cond1} when $\partial\Delta/\partial\eta
=0$.

At first order, one must solve
\begin{equation*}
\delta D_1 (a, p_a, \mu_{\bm{k}}, \pi_{\bm{k}}) = 
\delta D_1 (\tilde{a}, \tilde{p}_a, \tilde{\mu}_{\bm{k}},
\tilde{\pi}_{\bm{k}})
\end{equation*}
and
\begin{equation*}
\delta D_2 (a, p_a, \mu_{\bm{k}}, \pi_{\bm{k}}) = 
\delta D_2 (\tilde{a}, \tilde{p}_a, \tilde{\mu}_{\bm{k}},
\tilde{\pi}_{\bm{k}})
\end{equation*}
in order to determine the clock-transformed perturbation
variables. Explicitly, using \eqref{deltaD}, one gets
\begin{equation}
\begin{gathered}
\sqrt{k} \sin\Omega_k\,\mu_{\bm{k}} -
\frac{\cos\Omega_k}{\sqrt{k}}\,\pi_{\bm{k}} = \sqrt{k}
\sin\tilde{\Omega}_k\,\tilde{\mu}_{\bm{k}} -
\frac{\cos\tilde{\Omega}_k}{\sqrt{k}}\,\tilde{\pi}_{\bm{k}},\\ \sqrt{k}
\cos\Omega_k \,\mu_{\bm{k}} +
\frac{\sin\Omega_k}{\sqrt{k}}\,\pi_{\bm{k}} = \sqrt{k}
\cos\tilde{\Omega}_k\,\tilde{\mu}_{\bm{k}} +
\frac{\sin\tilde{\Omega}_k}{\sqrt{k}}\,\tilde{\pi}_{\bm{k}} ,
\label{tildemupi}
\end{gathered}
\end{equation}
where $\tilde{\Omega}_k=\frac12 k (\tilde{\eta} +
\tilde{a}/\tilde{p}_a) = \Omega_k + k\Delta$. The above algebraic
equations \eqref{tildemupi} can easily be inverted to yield the new
canonical perturbation variables, namely
\begin{equation}
\left(
\begin{array}{c}
\tilde{\mu}_{\bm{k}} \\ \\
\displaystyle\frac{\tilde{\pi}_{\bm{k}}}{k}
\end{array}
\right)
=
\left(
\begin{array}{cc}
\cos k\Delta & -\sin k\Delta \\ \\
\sin k\Delta &  \cos k\Delta
\end{array}
\right)
\left(
\begin{array}{c}
\mu_{\bm{k}} \\ \\ 
\displaystyle\frac{\pi_{\bm{k}}}{k}
\end{array}
\right).
\label{funfor}
\end{equation}
It is important to note that the above are classical relations between
canonical variables belonging to distinct canonical frameworks based
on distinct internal clocks. Although they are canonically
inequivalent, these two frameworks generate the same physical dynamics
of the system, which is required by the principle of relativity.

In general, clock transformations involve modifying temporal
relationships between events belonging also to different
spacetimes. This aspect of clock transformations is not reflected in
the lapse function $\tilde{N}$ of the new clock, which expresses the
temporal relationship between points within a single spacetime. The
clock transformations described in our framework however, do preserve
the foliation of cosmological spacetimes consisting of homogeneous
spatial leaves with small perturbations. Given that the initial clock
$\eta$ corresponds to the conformal time, the new lapse function of
the background foliation implied by the new clock $\tilde{\eta}$ reads
$$\tilde{N}=\frac{a}{1 + p_a \frac{\partial \Delta}{\partial a}}>0,$$
where ${\partial \Delta}/{\partial \eta}=0$ was assumed. Note that
considering a delay function satisfying $1 + p_a \partial
\Delta/\partial a =a$, one recovers the cosmic time with $\tilde N=1$.

\section{Quantization}
\label{sec4}

Having completed the classical treatment of our system, we now move to
the investigation of the possible differences between the respective
quantum dynamics obtained from the quantization of these two different
frameworks.

\subsection{Semiclassical background}

Since, by definition, the scale factor is positive definite ($a>0$),
one needs to quantize our previous system on the half line. Although
the position operator $\widehat{Q}=a$ is self-adjoint on the half
line, this is not the case for the momentum operator $\widehat{P} = i
\hbar\partial_a$, so we use instead the symmetric dilation
operator, $$\widehat{D} =\{ \widehat{P},\widehat{Q}\} = \frac12 \left(
\widehat{P} \widehat{Q} + \widehat{Q} \widehat{P}\right) = \frac12
i\hbar \left( a\partial_a + \partial_a a\right).$$ Classically the
dilation variable is $d = a p_a$, so that the Hamiltonian, expressed
in terms of $d$, is $H^{(0)} = \frac12 p_a^2 = \frac12 d^2/a^2$, and
one can define its quantum counterpart as a symmetric ordering of
$\frac12 \widehat{Q}^{-2} \widehat{D}^2$.  Expanding on the basis
($\widehat{Q},\widehat{P}$), this yields
$$ \widehat{H}^{(0)} = - \frac{1}{2}\frac{\partial^2}{\partial a^2} +
\frac{\hbar^2K}{a^2},
$$
where the value of $K>0$ depends on the ordering; fixing one ordering
such that $K>\frac34$ ensures $\widehat{H}^{(0)}$ is self-adjoint on
the half line~\cite{Vilenkin:1987kf}.

We can find some approximate solutions to the Schr\"odinger equation
with a family of coherent states (see, e.g.,
Refs.~\cite{Klauder:2015ifa,Klauder:2012vh, Martin:2021dbz} for the specific case under study here). We choose the coherent states
to read
\begin{equation}
|a(\eta),p_a(\eta)\rangle = \ex^{i p_a(\eta) \widehat{Q}/\hbar}
\ex^{-i \ln [a(\eta)] \widehat{D}/\hbar}|\xi\rangle,
\label{CohSta}
\end{equation}
where $|\xi\rangle$ is such that the expectation values of
$\widehat{Q}$ and $\widehat{P}$ in $|a(\eta),p_a(\eta)\rangle$ are,
respectively, $a(\eta)$ and $p_a(\eta)$, and otherwise arbitrary (see,
however, Ref.~\cite{Bergeron:2023zzo}).

\begin{figure}[t]
\hskip-5mm
  \includegraphics[width=0.5\textwidth]{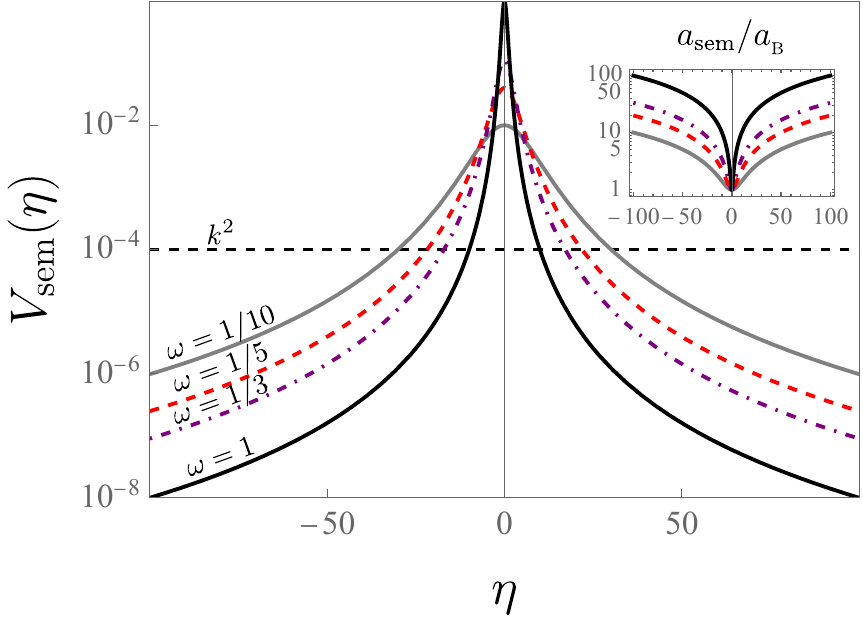}

  \caption{The semiclassical potential $V_\text{sem}$ given by
    Eq.~\eqref{Vsem} as a function of the conformal time $\eta$ for
    various values of the inverse bounce duration $\omega$. The
    potential has to be compared with the relevant value of $k^2$
    ($k=0.01)$, indicated as a straight line. The corresponding scale
    factor time evolution is shown in the inset.}

  \label{VsemFig}
\end{figure}

The dynamics confined to the coherent states can be deduced from the
quantum action
\begin{equation}
\mathcal{S}_\textsc{q} =\int \left\{ a'(\eta) p_a(\eta) -
H_\text{sem}\left[a(\eta),p_a(\eta)\right] \right\}\dd \eta,
\end{equation}
with the semiclassical Hamiltonian given by
\begin{equation}
 H_\text{sem} =\langle a,p_a| \widehat{H}^{(0)} |a,p_a\rangle,
\end{equation}
from which one derives the ordinary Hamilton equations,
\begin{equation}
a' = \frac{\partial {H}_\text{sem}}{\partial p_a} \quad \text{and}
\quad p'_a = - \frac{\partial {H}_\text{sem}}{\partial
  a}. \label{eomsem}
\end{equation}
We find that the semiclassical background Hamiltonian reads
\cite{Martin:2021dbz}
\begin{equation}
H_\text{sem}=\frac12 \left( p_a^2 + 
\frac{\hbar^2\mathfrak{K}}{a^2}\right),
\label{HK}
\end{equation}
where the new constant $\mathfrak{K}$ is positive ($\mathfrak{K}>0$).
Its specific value is related with both $K$ and the fiducial state
$|\xi\rangle$. We find the solution to \eqref{eomsem} to read $a^2(\eta) = a_0 + a_1 \eta + a_2 \eta^2$, with $a_0 a_2
-a_1^2/4 =\hbar^2\mathfrak{K} >0$, so that the equation $a(\eta) =0$
has no longer any real solution; the singularity is indeed quantum
mechanically avoided. Choosing the origin of time such that $a'=0$ for
$\eta=0$ permits us to rewrite this solution in full generality as
\begin{subequations}
\begin{align}
a(\eta) & = a_\textsc{b} \sqrt{1+(\omega \eta)^2},
\label{aeta}\\
p_a(\eta) & = \frac{a_\textsc{b} \omega^2 \eta}
{\sqrt{1+(\omega\eta)^2}},
\label{paeta}
\end{align}
\label{solsem}
\end{subequations}
where $a_\textsc{b}^4 \omega^2 = \hbar^2\mathfrak{K}$, which in turn
implies $H_\text{sem} = \frac12 a_\textsc{b}^2 \omega^2
  =\frac12 \hbar \sqrt{\frak{K}} \omega = \hbar^2 \frak{K}/(2
  a_\textsc{b}^2)$; it is clear that the model contains one and only
one free parameter, namely $\frak{K}$.  From now on, we assume that
the background evolution is given by Eqs.~\eqref{solsem}: this means
the semiclassical potential
\begin{equation}
V_\text{sem} = \frac{a''}{a} = \frac{\hbar^2 \frak{K}}{a^4} =\left[
  \frac{\omega}{1+\left(\omega\eta\right)^2} \right]^2,
\label{Vsem}
\end{equation}
shown in Fig.~\ref{VsemFig}, never vanishes except in the large scale
factor limit ($a\gg 1 \Longrightarrow \eta \gg \omega^{-1}$). This is
appropriate as this is also the classical limit for which $a''\to
0$. A classical radiation-dominated universe begins or ends with a
singularity and produces no gravitational waves, whereas our quantum
radiation-dominated universe naturally connects the contracting and
expanding phases through a bounce, which is subsequently responsible
for a nonvacuum spectrum of tensor perturbations, to which we now
turn.

\subsection{Quantum perturbations}

For a given mode $\bm{k}$, the Hamiltonian $H^{(2)}_{\bm{k}}$, given
by Eq.~\eqref{H2k}, is easily quantized using the usual
prescriptions. We assume that the background follows the semiclassical
approximation described above, so that the potential for the
perturbation is given by $V_\text{sem}$ [Eq.~\eqref{Vsem}]. The basic
variables are replaced by a set of operators
\begin{equation}
\begin{gathered}
\mu_{\bm{k}} \mapsto \widehat{\mu}_{\bm{k}} =
\sqrt{\frac{\hbar}{2}}\left[\widehat{a}_{\bm{k}} \mu^*_k(\eta)+
  \widehat{a}^{\dagger}_{-\bm{k}} \mu_k(\eta)\right],\\ \pi_{\bm{k}}
\mapsto \widehat{\pi}_{\bm{k}} = \sqrt{\frac{\hbar}{2}}
\left[\widehat{a}_{\bm{k}} \mu^{*\prime}_k(\eta)+
  \widehat{a}^{\dagger}_{-\bm{k}} \mu'_k(\eta)\right],
\end{gathered}
\label{operators}
\end{equation}
where we assume the Wronskian normalization condition $\mu'_k \mu^*_k
- \mu_k \mu^{*\prime}_k = 2i$ for the complex mode functions
$\mu_k$. The creation $\widehat{a}^\dagger_{\bm{k}}$ and annihilation
$\widehat{a}_{\bm{k}}$ operators satisfy the commutation relations
$\left[ \widehat{a}_{\bm{k}}, \widehat{a}^\dagger_{\bm{p}} \right] =
\delta_{\bm{k}, \bm{p}}$ stemming from the canonical ones between the
field operators $\left[\widehat{\mu}_{\bm{k}}, \widehat{\pi}_{-\bm{p}}
  \right] = i\hbar \delta_{\bm{k}, \bm{p}}$.

\begin{figure}[th]
  \includegraphics[width=0.45\textwidth]{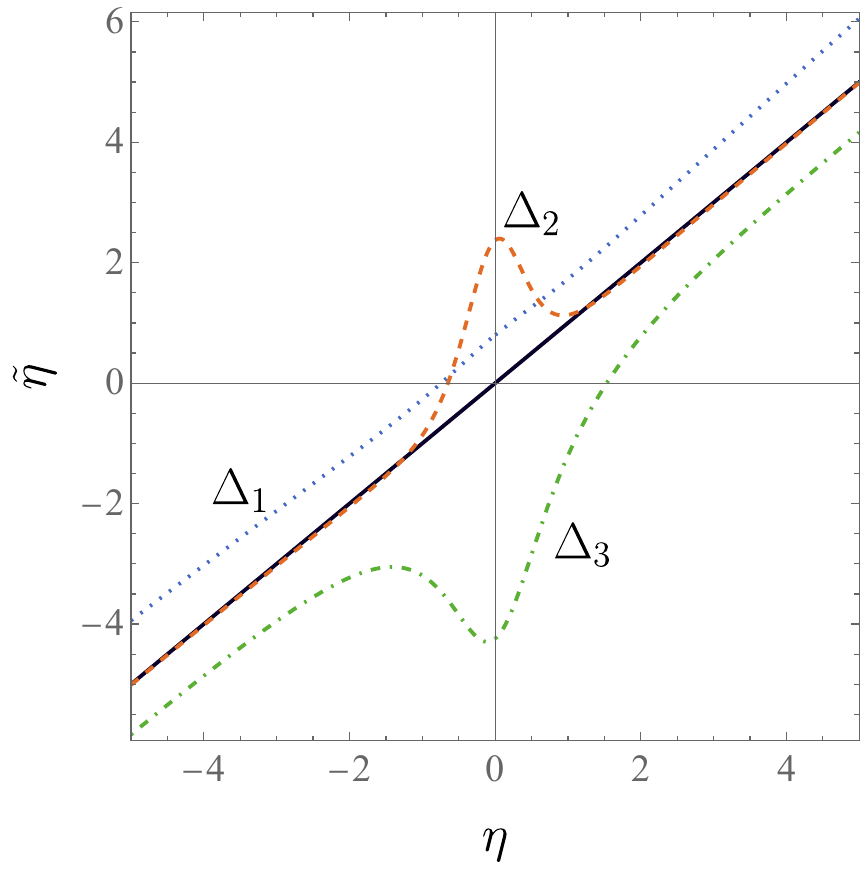}

  \caption{The new time $\tilde{\eta}$ as a function of the original
    one $\eta$ for three different shapes of delay functions
    $\Delta_1$, $\Delta_2$, and $\Delta_3$ defined through
    Eq.~\eqref{delfun} along the original fixed bouncing trajectory
    \eqref{solsem}.  The parameters are chosen as $A=B=D=1$, $C=4$,
    and $E=2$ for $\Delta_1$, while we set $A=2$, $B=0.2$, $C=0.5$,
    $D=3$, and $E=4$ for $\Delta_2$, and finally the set $A=-1$,
    $B=C=1$, $D=0.5$, and $E=3$ defines $\Delta_3$.}

  \label{deltas}
\end{figure}

In the Heisenberg picture, the equations of motion take the form
$$ i\hbar\frac{\dd \widehat{\mu}_{\bm{k}}}{\dd\eta} = \left[
  H^{(2)}_{\bm{k}}, \widehat{\mu}_{\bm{k}}\right] \quad \hbox{and}
\quad i\hbar\frac{\dd \widehat{\pi}_{\bm{k}}}{\dd\eta} = \left[
  H^{(2)}_{\bm{k}}, \widehat{\pi}_{\bm{k}}\right],
$$
which imply that the mode function $\mu_k(\eta)$
satisfies
\begin{equation}
\frac{\ud^2\mu_k}{\ud\eta^2}+\left(k^2 -
\frac{\hbar^2\mathfrak{K}}{a^4}\right)\mu_k=0,
\label{mode1}
\end{equation}
where $a(\eta)$ is given by the semiclassical solution
\eqref{aeta}. Using \eqref{Vsem}, this transforms into
\begin{equation}
\frac{\ud^2\mu_k}{\ud\eta^2}+\left\{ k^2 -
\left[ \frac{\omega}{1+\left(\omega\eta\right)^2} \right]^2
\right\}
\mu_k=0,
\label{modemu}
\end{equation}
which can be integrated numerically if initial conditions are
provided: we assume that far in the contracting branch, with
$\eta_\text{ini} <0$ and $V_\text{sem}(\eta_\text{ini}) \ll k^2$,
there was no gravitational wave, so the field was in a vacuum
state. This implies the mode function satisfies $\mu_k
(\eta_\text{ini}) = \ex^{-ik\eta_\text{ini}}/\sqrt{2k}$ and $\mu'_k
(\eta_\text{ini}) =-i\sqrt{k/2} \, \ex^{-ik\eta_\text{ini}}$.

\section{Quantum ``clocks''}
\label{sec5}

In what follows, we study the effect of clocks on the quantum and
semiclassical dynamics of selected dynamical variables. First, we
obtain the dynamical trajectories in the reduced phase space
$(a,{p},{\mu}_k,\pi_k)$ that is associated with the initial clock
${\eta}$; note that, from that point on, since there is no risk of
confusion, we shall replace what was previously denoted as $p_a$
simply by $p$. Next, we choose a set of delay functions $\Delta(a,p)$
to define new clocks $\tilde{\eta}$ and obtain the new reduced phase
spaces $(\tilde{a}, \tilde{p},\tilde{\mu}_k,\tilde{\pi}_k)$ associated
with the new clocks. Then, we make use of Eqs.~\eqref{CT},
\eqref{CTbckg}, and \eqref{funfor} to transport the dynamical
trajectories to these new phase spaces. We assume that the latter
admit a unique physical interpretation, and so the trajectories can be
meaningfully compared in these new variables. In other words, there
are many clocks denoted by ${\eta}$ and only one denoted by
$\tilde{\eta}$. Note that for $\Delta=0$ the clocks ${\eta}$ and
$\tilde{\eta}$ coincide. For this case, we assume that ${\eta}$ and
$(a,{p},{\mu}_k,\pi_k)$ are the variables of Sec. \ref{sec2}, which
sets the physical meaning of the phase space $(\tilde{a},
\tilde{p},\tilde{\mu}_k,\tilde{\pi}_k)$ and the clock $\tilde{\eta}$.

\begin{figure}[t]
\includegraphics[width=0.45\textwidth]{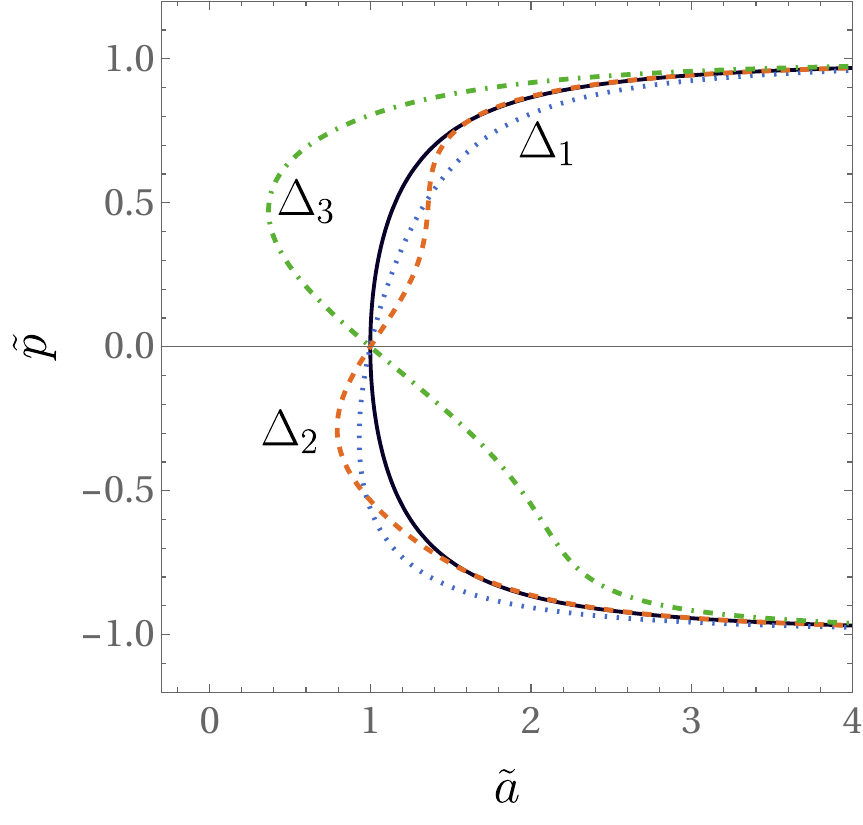}

\caption{Semiclassical trajectories obtained in different clocks and
  mapped into the initial reduced phase space $(\tilde{a},\tilde{p})$
  to compare with the original trajectory represented by the full
  black line.}

\label{qpbackground}
\end{figure}

\begin{figure}[t]
  \includegraphics[width=0.45\textwidth]{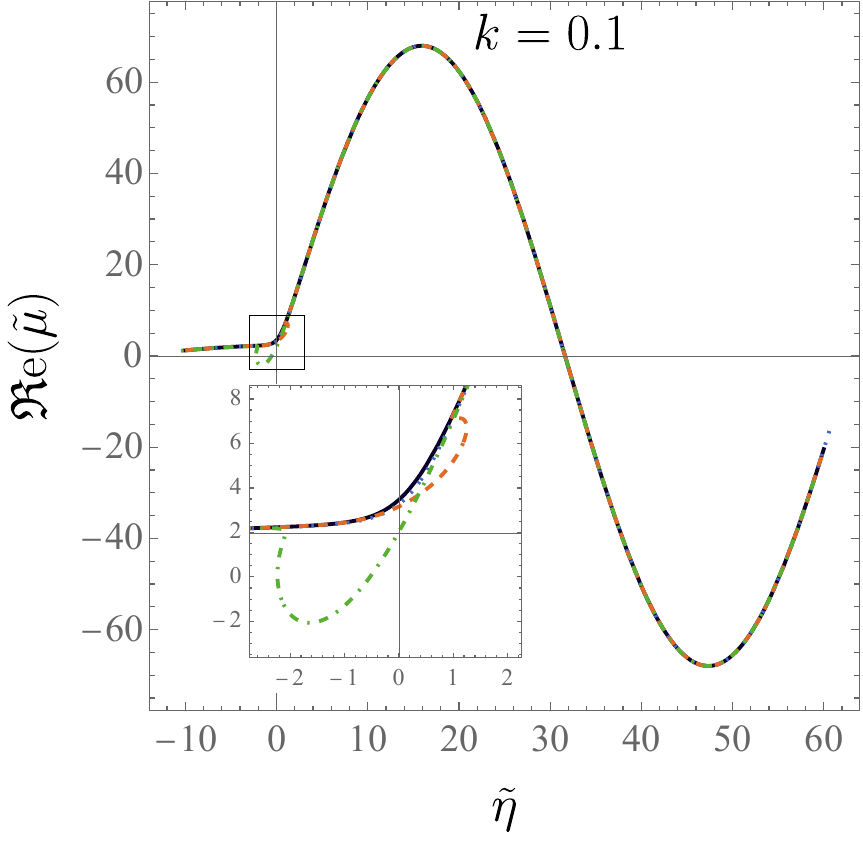}
  \includegraphics[width=0.45\textwidth]{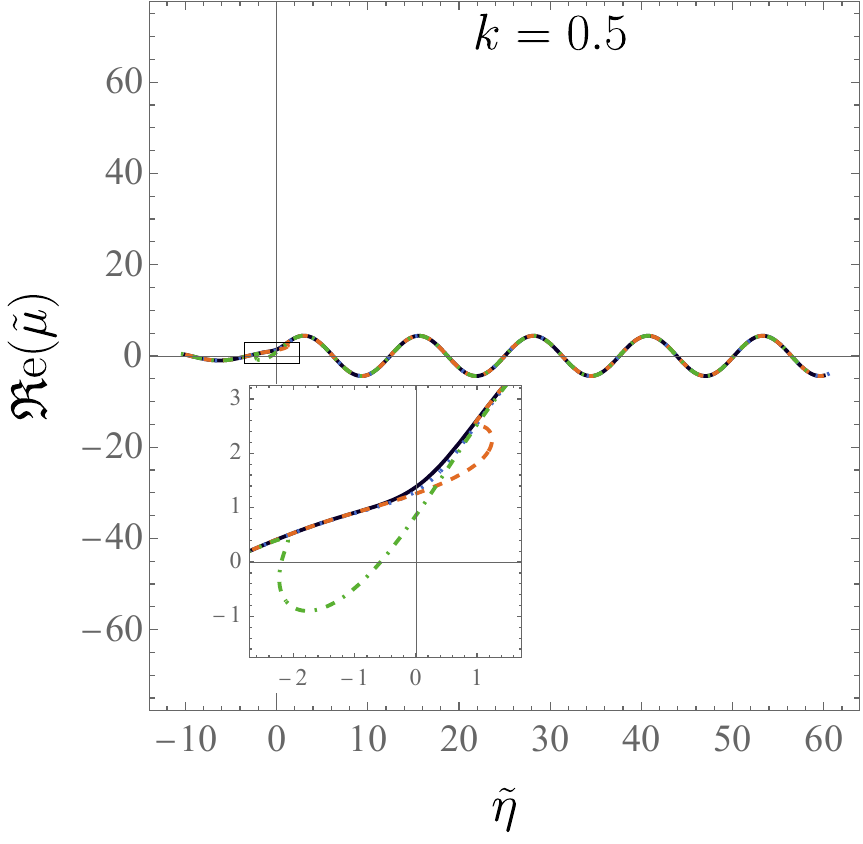}

  \caption{Evolution of the primordial gravity wave
    $\mathfrak{R}\mathrm{e}(\tilde{\mu})$ for two different wave
    numbers, $k=0.1$ (top) and $k=0.5$ (bottom), and for different
    clocks based on the first class of delay function, $\Delta_1$,
    $\Delta_2$, and $\Delta_3$, represented by the dotted blue line,
    dashed red line, and dash-dotted green line, respectively. The
    original trajectory is represented by the full black line. In
    Fig. \ref{mu-k-clock2}, the same plot for the second class of
    delay function is depicted to show how the choice of delay
    function affects the time of convergence.}

  \label{mu-k-clock}
\end{figure}

\begin{figure}[ht]
   \includegraphics[width=0.45\textwidth]{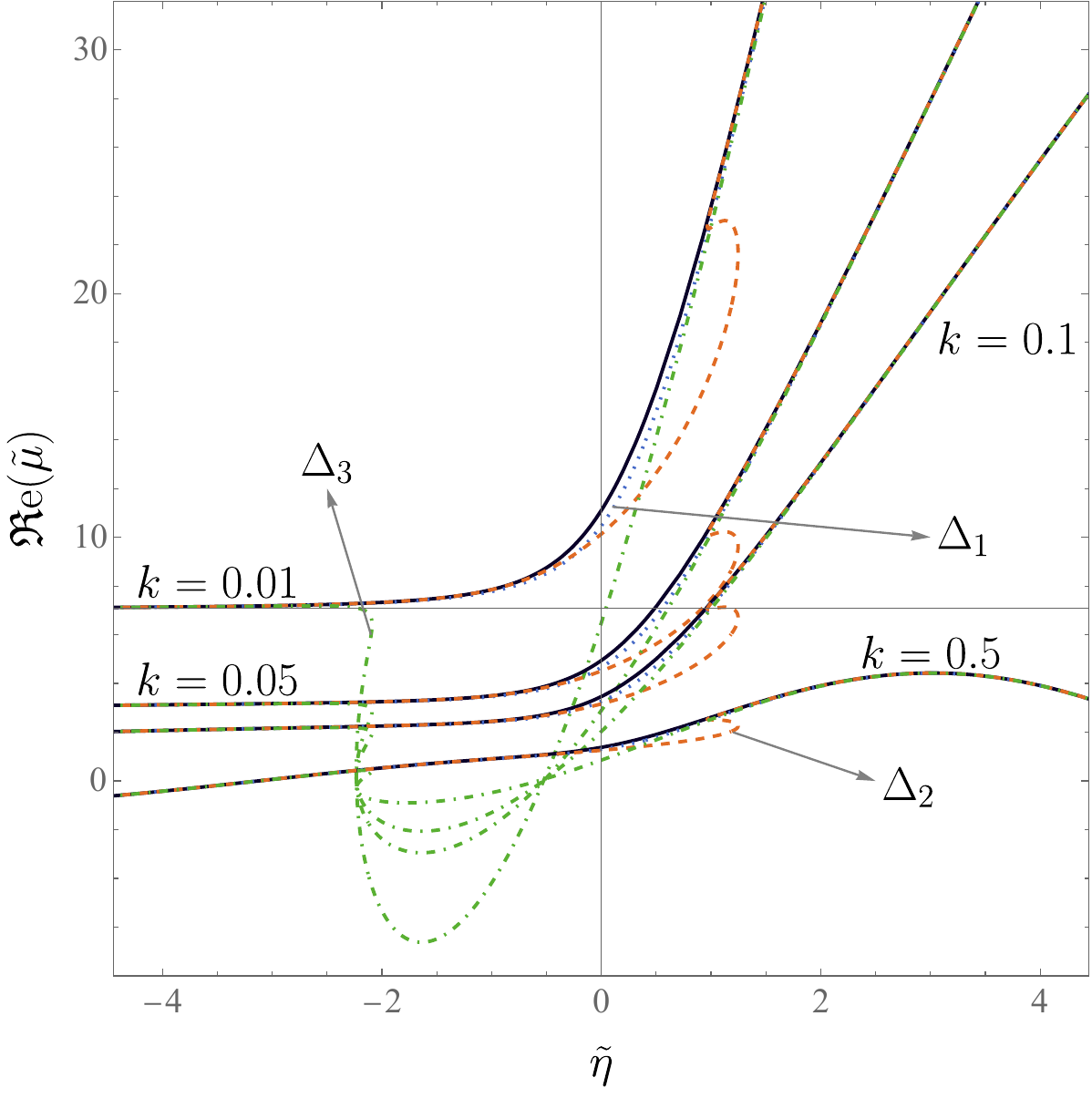}

   \caption{Evolution of the primordial gravity wave
     $\mathfrak{R}\mathrm{e}(\tilde{\mu}_k)$ plotted for four
     different wave number $k$ values. For each fixed $k$ we changed
     the clock considering the family of delay functions $\Delta$,
     whose value is the same as in Fig. \ref{qpbackground}.}
   
 \label{mu-all-k}
\end{figure}

\subsection{Clock choices and background}

In order to illustrate the clock choice issue, we consider a family of
delay functions, namely
\begin{equation}\label{delfun}
\Delta(a,p)=A \frac{a^B}{(a+C)^D}\frac{\sin(E p)}{p},
\end{equation}
where $A$, $B$, $C$, $D$, and $E$ are arbitrary coefficients, whose
values are limited to ensure that the conditions presented in
Sec.~\ref{clocktra} hold. In the Appendix, we consider another set of
acceptable delay functions to show that our conclusions are not
restricted to the choice \eqref{delfun}.

A few clocks corresponding to the delay function $\Delta(a,p)$ are
represented along a semiclassical dynamical trajectory for different
choices of the free parameters in Fig.~\ref{deltas}. It shows that,
contrary to the classical case where the condition \eqref{Cond1}
holds, the new clocks, in general, are no longer monotonic due to
quantum corrections.

Applying the clock transformation of Fig.~\ref{deltas} to the
background solution \eqref{solsem} yields Fig.~\ref{qpbackground} once
mapped into the reduced phase space, with the original trajectory
superimposed for comparison.  All the trajectories originate in the
same classical regime at large $\tilde{a}$ and negative $\tilde{p}$,
i.e., at a time at which the universe is large and contracting. Close
to the $\tilde{a}=0$ boundary, where the quantum behavior dominates,
they all somehow bounce in the variables $\tilde{a}$ and $\tilde{p}$,
diverging from one another and providing different accounts of the
bounce. Finally, they reach the region of large $\tilde{a}$ and
positive $\tilde{p}$ where they converge again to the unique classical
behavior representing a large and expanding universe.

Possible differences between the trajectories include the values of
$\tilde{a}$ and $\tilde{p}$ at which the bounce occurs, the level of
asymmetry between contracting and expanding branches, or even the
number of bounces. These semiclassical trajectories illustrate the
nonunitary relation between different clocks.  Nevertheless, they all
originate from a unique contracting classical universe and end toward
a similarly unique expanding classical universe. Therefore, the
semiclassical trajectories in different clocks yield the same outcome
for large and classical universes.  Notice that the trajectories'
convergence before and after the bounce can be delayed as much as one
wants by making use of appropriate delay functions, such as that
discussed in the Appendix, i.e., Eq.~\eqref{delfun2}, whose effects on
both background and perturbation trajectories can be seen in the
Appendix.

Let us now move to the perturbation of these homogeneous solutions and
compare the different evolution that can result from using different
clocks.

\subsection{Clocks and perturbations}

In Fig.~\ref{mu-k-clock}, we plot the dynamics of the real part of the
perturbation variable $\tilde\mu_k$ against the delayed time
$\tilde{\eta}$ for the three different functions of Eq.~\eqref{delfun}
displayed in Fig.~\ref{deltas} and for two values of the comoving
wave number $k$. The figure illustrates our general finding that the
absolute clock effect is more or less equally strong and lasting
roughly equally long for all wavelength perturbations. This is shown
more convincingly in Fig.~\ref{mu-all-k}, in which the evolution of
four different modes is shown as a close-up in the quantum-dominated
bouncing region. This means that the larger the wavelength of the
perturbation, the larger the relative clock effect, and the longer it
lasts in units of its oscillation period. Thus, the clock effect is
more important for phenomena occurring at small timescales and over
short distances.  Moreover, the evolving amplitude $\tilde\mu_k$, in
general, is not a function of the clock $\tilde{\eta}$ due to quantum
effects that disrupt the monotonicity relation between quantized
clocks.

Given that both the background and the perturbation modes evolve in
such a way as to reach a unique configuration, the primordial
gravity-wave amplitude $\tilde\mu_k/\tilde{a}$, which is the quantity
one expects to measure in practice~\cite{Micheli:2022tld}, also
converges to a unique solution, making the model predictive.

All the plots above illustrate the nonunitary relation between
different clocks, as well as the spoiling of the clock monotonicity at
the quantum level, which is illustrated in Fig.~\ref{deltas}.
Nevertheless, similar to the semiclassical background trajectories,
the perturbation variable $\mathfrak{R}\mathrm{e}(\tilde{\mu})$
visibly converges to a unique classical solution from a well-defined
asymptotic past initial condition to the asymptotic future. Therefore,
one can safely extend the background conclusion to the perturbations:
the time development of the mode $\mathfrak{R}\mathrm{e}(\tilde{\mu})$
using different clocks yields the same predictions in the large and
classical universe regime.  The delay of the convergence due to
different choices of delay functions can be seen in
Fig.~\ref{mu-all-k}.

\begin{figure}[h]
  \includegraphics[width=0.48\textwidth]{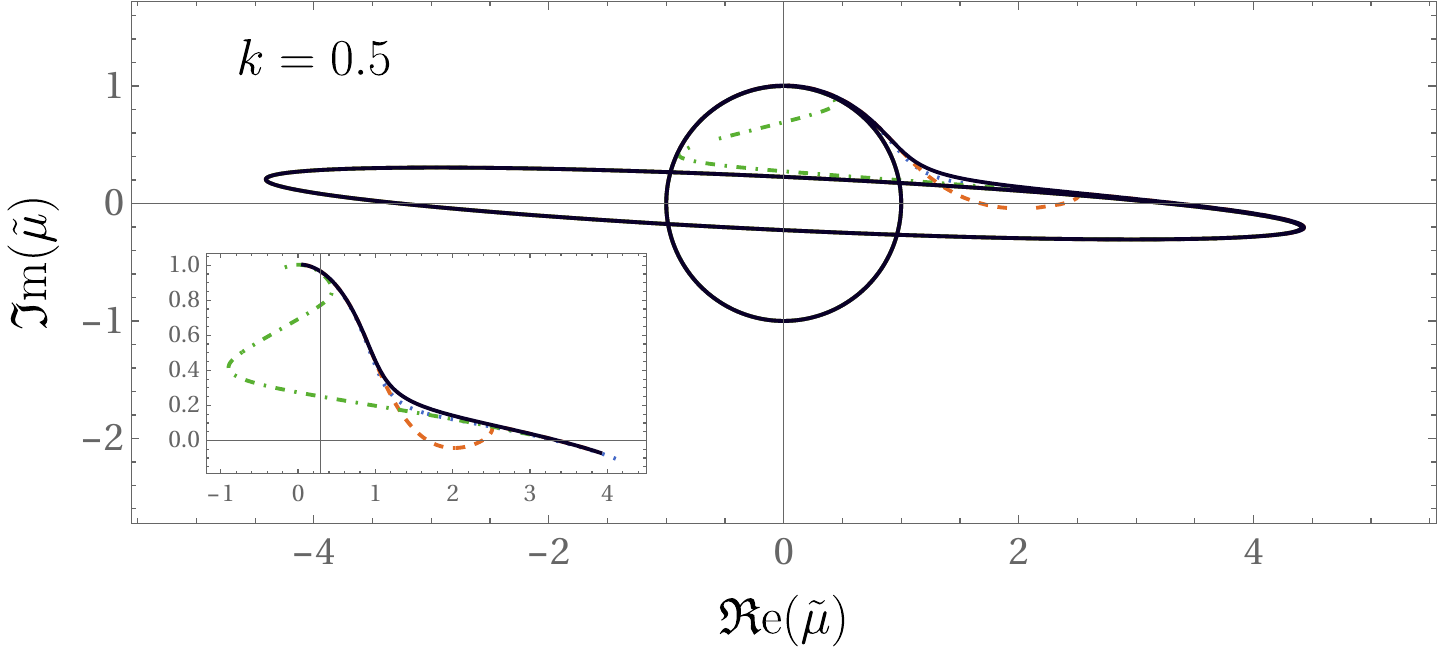}

  \caption{Evolution of the real versus imaginary part of
    $\tilde{\mu}_k$ for a wave number $k=0.5$ and a bounce parameter
    $\omega=1$.  The initial circle represents the initial vacuum
    state of the perturbation, while the ellipse shows the final
    squeezed state, which happens, in the case at hand, to have a
    slight phase shift with respect to the real axis. The transition
    between these two asymptotic cases differs for the different delay
    functions $\Delta_1$, $\Delta_2$, and $\Delta_3$, whose
    trajectories are represented by the dotted blue line, dashed red
    line, and dash-dotted green line, respectively, the original
    trajectory being represented by the full black line.}

  \label{RealIMmu}
\end{figure}

As a final illustration of the perturbation behavior through the
quantum bounce, we find it useful to inspect the phase space
trajectories in the plane
$[\mathfrak{R}\mathrm{e}(\tilde\mu_k),\mathfrak{I}\mathrm{m}(\tilde\mu_k)]$
as is displayed in Fig. \ref{RealIMmu}. The initial vacuum state is
represented by a circle that is squeezed into an ellipse during the
contraction and bounce, squeezing that represents the amplification of
the amplitude of the perturbation. From the point of view of the time
problem, the initial circle and the final ellipse, respectively,
represent the asymptotic past and future of the amplitude: from the
point of view of physical prediction, the indeterminacy occurring near
the bounce, as may develop through various different times disappears
in the asymptotic regimes, so that the existence of a classical
approximation in our trajectory approach ensures the standard
procedure of treating the perturbations leads to physically
meaningfull predictions.

\section{Discussion and perspectives}\label{sec6}

In this work, we explored the time problem in the framework of quantum
fields on quantum spacetimes. We considered the specific example of
primordial gravitational waves propagating through a bouncing quantum
Friedmann universe. We pointed to several features that we believe to
be universal for such models.

First, we showed that the dynamical variables, such as the scale
factor or the amplitude of a gravitational wave, obtained from
different internal clocks, evolve differently when compared in a
clock-independent manner. Second, these expectation values (background
evolution) and mode functions of operators (perturbations),
irrespective of the clock chosen, converge to a unique evolution for
large classically behaving universes. This is the phase space domain
in which unambiguous predictions can be made. Third, for different
clocks, the dynamics converges to the classical behavior at different
times. In principle, there is no restriction on how far from the
bounce the system must be in order to display the classical
behavior. In practice, however, all the clocks considered were found
to converge very quickly, allowing for unambiguous predictions shortly
after the bounce.

Based on the above findings, we postulate that the physical
predictions are only those predictions provided by any clock, which
are not altered upon the clock's transformation. The fact that for
large universes the semiclassical background dynamics and the quantum
perturbation dynamics do not depend on the clock implies the
following: {\it Despite the fact that the dynamical variables are not
  Dirac observables, they provide physical predictions for large
  universes, which is precisely the regime in which we observe the
  actual Universe}.

Note, however, that the word ``large'' is never precisely defined. One
could expect that, at least in principle, some clocks require times
larger than the present age of the Universe to converge to the
classical behavior. This, however, poses no problem to our
interpretation, as we simply exclude such clocks and retain only those
that behave classically in the domain for which we make
predictions. This may seem arbitrary and unjustified. We must,
however, remember that, as a matter of fact, any semiclassical
description of ordinary quantum mechanics is necessarily restricted to
a limited set of observables, usually the simple ones, while more
compound observables often display classically incompatible behavior
(e.g., $\langle x \rangle^2 \not= \langle x^2 \rangle$).  For similar
reasons, we are allowed to choose only those clocks in which the
dynamics of the relevant observables is classically consistent.

On the one hand, we proved that the evolution of the expectation
values of some observables constitute {\it physical} predictions of
quantum cosmological models. On the other hand, the expectation values
are not all that is measured in the large Universe. In other words,
not all objects are classical in the large Universe. For instance, the
position of an electron is a dynamical variable that can be measured
in a laboratory. So, could the outcomes of such a measurement also be
unambiguously predicted by a quantum cosmological model? The answer is
affirmative. Note that the mode function $\mu_k$, whose dynamics
becomes unambiguous in a large universe, determines the evolution of the
operator $\widehat{\mu}_k$ via Eq.~\eqref{operators}. This implies
that the Heisenberg equation of motion encoded in Eq.~\eqref{modemu}
becomes unambiguous too. Obviously, the evolution of perturbation in
the Schr\"odinger picture must consequently become unique as
well. Hence, ordinary quantum mechanics of perturbation modes is
recovered in a large universe. These conclusions must also apply to
electrons and, in general, to all nongravitational degrees of freedom.

To better understand the origin of the emergence of ordinary quantum
mechanics, notice that any clock transformation \eqref{CT} involves,
by definition, only background variables. If the latter behave
classically, the clock transformation is completely classical and
amounts to a mere (in general, nonlinear) change of units of time. In
Ref.~\cite{Malkiewicz:2017cuw}, it was demonstrated that the
relational dynamics of {\it a quantum variable in a classical clock}
is unambiguous in the sense that switching to another classical clock
does not induce any clock effect.

Let us put to test our approach and our result by addressing a set of
questions that were proposed in Ref.~\cite{Isham:1992ms} for assessing
the completeness of any potential solution to the time problem.

\begin{enumerate}[label=(\arabic*)]

  \item How should the notion of time be re-introduced into the
    quantum theory of gravity?

\hskip2mm Our approach relies on evolving internal variables called
clocks. We express the dynamics of the dynamical variables in terms of
these clocks.

\item In particular, should attempts to identify time be made at the
  classical level, i.e., before quantization, or should the theory be
  quantized first?

\hskip2mm In our approach, we first reduce the Hamiltonian formalism
based on a selected clock, then we quantize the reduced formalism as
if the clock was an external and absolute time. However, it is neither
external nor absolute. The instantaneous value of the clock determines
the instantaneous physical state of the system. Switching to another
clock entails a change in the physical interpretation of the clock and
the entire state of the system.

\item Can ``time'' still be regarded as a fundamental concept in a
  quantum theory of gravity, or is its status purely phenomenological?
  [...]

\renewcommand{\thefigure}{A\arabic{figure}}
\setcounter{figure}{0}

\begin{figure}[b]
  \includegraphics[width=0.45\textwidth]{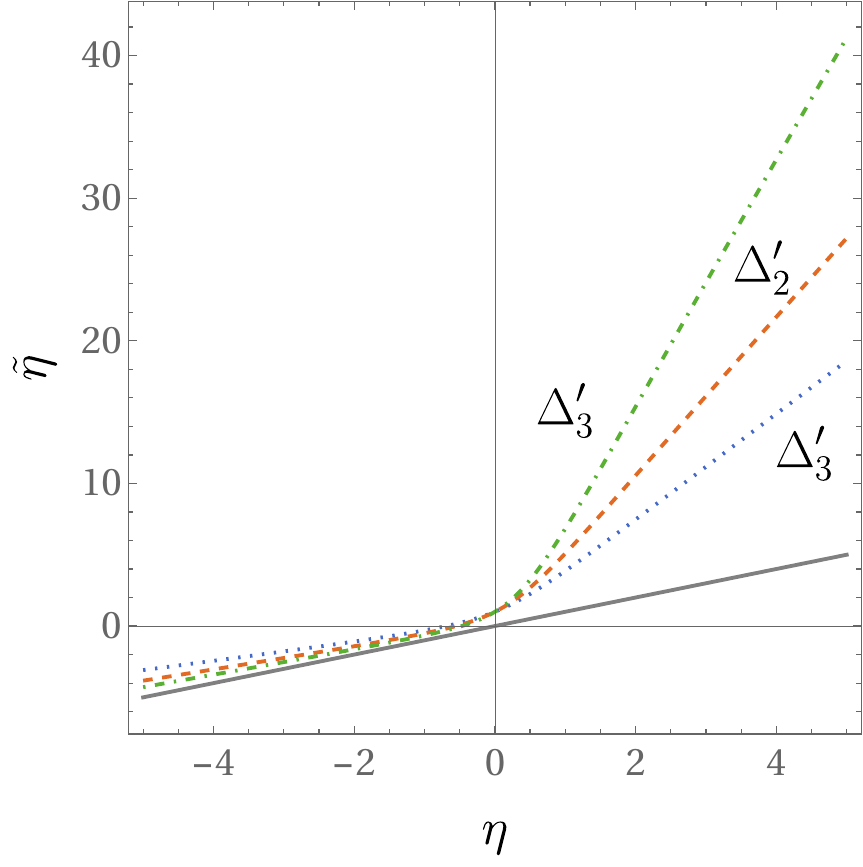}

  \caption{Changes in the time variable $\eta$ for the second family
    of delay functions $\Delta'_1$, $\Delta'_2$, and $\Delta'_3$ given
    by Eq.~\eqref{delfun2} along a fixed bouncing trajectory, with
    parameters chosen such that $\Delta_1=a e^{p_a}$, $\Delta_2=a
    e^{3p_a/2}$, and $\Delta_3=a e^{2p_a}$. }

  \label{deltasprime}
\end{figure}

\begin{figure}[t]
  \includegraphics[width=0.45\textwidth]{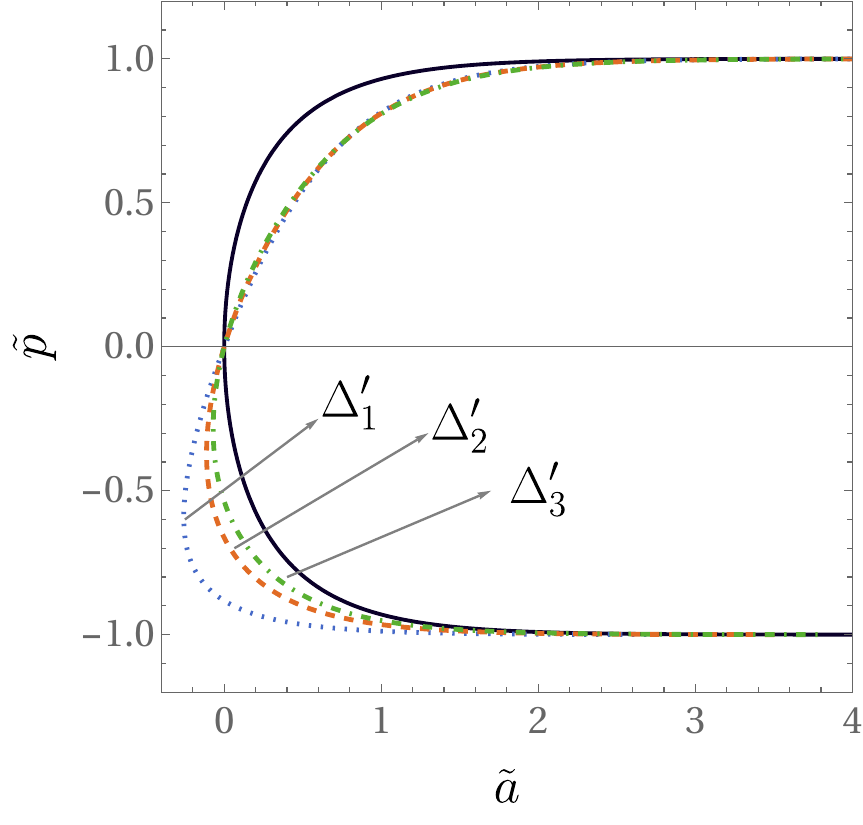}

  \caption{Semiclassical trajectories mapped into the initial reduced
    phase space $(a, p)$ for the second class of delay function
    \eqref{delfun2}, with the same parameters as in
    Fig.~\ref{deltasprime}.}

  \label{qpdeltaprime}
\end{figure}

\begin{figure}
  \includegraphics[width=0.45\textwidth]{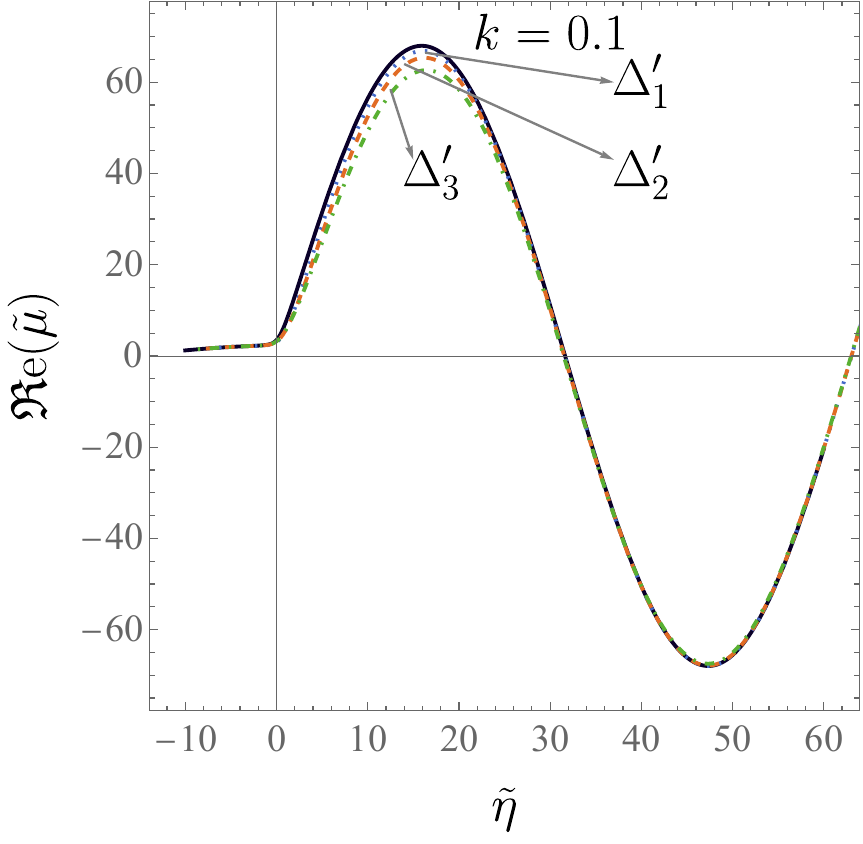}
  \includegraphics[width=0.45\textwidth]{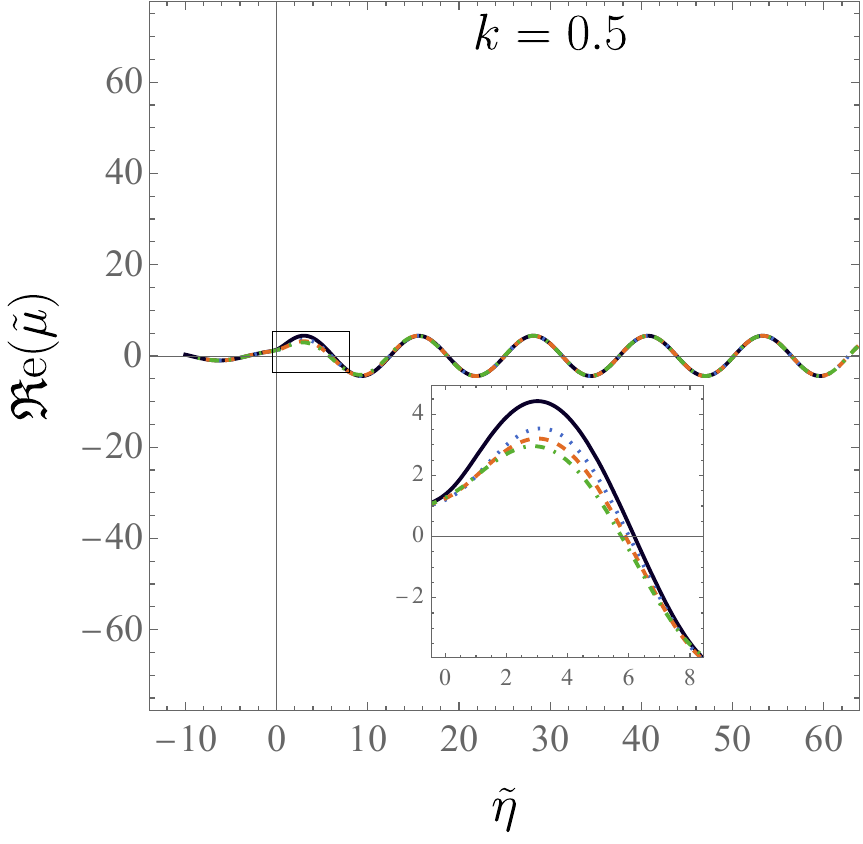}
  \caption{Evolution of the real part of the primordial gravity wave
    $\mathfrak{R}\mathrm{e}(\tilde{\mu})$ for two different wave
    numbers, $k=0.1$ and $k=0.5$, and for different clocks for the
    second class of delay functions, $\Delta'_1$, $\Delta'_2$, and
    $\Delta'_3$, respectively, represented by the dotted blue line,
    dashed red line, and dash-dotted green line. The original
    trajectory is represented by the full black line.}

  \label{mu-k-clock2}
\end{figure}

\hskip2mm In our approach, there is no fundamental time. The
fundamental concept is ``change'' or ``evolution,'' meaning we merely
need to assume that the $3+1$ split of the underlying geometry imposes
an ordered set of hypersurfaces. As we showed in this paper,
extracting dynamical predictions from such a formalism is a subtle
issue. The clocks serve as tools for deriving the predictions. Once a
class of clocks converges to a unique dynamics, any one of them can be
treated analogously and deserve the qualification of time, and any
quantum dynamical variable becomes described in them by a unique
Schr\"odinger equation. This is how ordinary quantum mechanics
emerges.

\item If time is only an approximate concept, how reliable is the
  rest of the quantum-mechanical formalism in those regimes where the
  normal notion of time is not applicable? In particular, how closely
  tied to the concept of time is the idea of probability? [...]

\hskip2mm The quantum-mechanical description in the regime where
different clocks exhibit different dynamics is an essential part of
our theory. It describes the deterministic evolution of the
system. However, this regime does not seem to allow for any meaningful
dynamical interpretation in terms of relational change.  Although we
have not explicitly addressed this question in the present work, our
approach permits one to do it.
\end{enumerate}

\bigskip

To conclude, one can mention that the chosen clock degrees of freedom,
although perfectly acceptable as such in the classical framework of
general relativity, are arguably not in the quantum regime. They do
not qualify as actual clocks since, along the quantum trajectory, they
yield a nonmonotonic change of time variable; in other words, they
provide different hypersurface orderings. This might be cured by
adding to the classical clock transformation \eqref{CT} a quantum term
that needs be identified. One may also argue that we are insisting
upon using a trajectory to define the background evolution, while some
might insist upon the fact that there is no such thing as a trajectory
in quantum mechanics.

In any case, it is interesting to note that, whichever of the
possibilities above happens to be valid, the critical point that is
made here is that, even though the quantum-dominated phase is indeed
ill-defined both at the background and perturbation levels from the
point of view of time development, the asymptotic regimes end up being
unique. As a result, setting well-motivated initial conditions in the
classical past, one gets unambiguous physical predictions for the
classical future in which we happen to perform the ensuing
measurements. In other words, we have shown that the lack of
predictability in the quantum regime does not exclude the fact that
the theory permits meaningful physical predictions that can be tested
with observations.

Finally, it is worth noting that there are alternative approaches that
do not involve promoting internal variables to clock status,
effectively avoiding the time problem.  For instance, in
Ref.~\cite{Brizuela_2016}, the Wentzel-Kramers-Brillouin approximation
to the background wave function is made, and the resultant trajectory
provides a well-defined cosmological background on which perturbations
propagate, without ever introducing the physical inner product at the
background level. An approach similar in spirit can be found in
Ref.~\cite{PhysRevD.67.063517}, which is based on coarse graining of
the background wave function, thereby removing short timescale
oscillations in the scale factor. The end result is similar to the
previous case and allows for an unambiguous effective trajectory in
the background variables along which the evolution of the
perturbations occur. Although the time problem discussed in the
present work is absent in these approaches, the cost is that of a
limited physical interpretation of the background wave function, for
which no notion of unitary dynamics is ever introduced. Consequently,
the quantum uncertainties in the physical background variables are not
well defined and thus their influence on the dynamics of the
perturbations is assumed negligible. Choosing an internal time entails
a prescription for calculating such uncertainties and permits one to
incorporate them in the dynamics of perturbations. The resulting clock
dependence of such a prescription leads to the questions addressed
here.

A unitary approach to quantum cosmology that aims at a
gauge-independent formulation was described in
Ref.~\cite{Barvinsky:1993jf}. This interesting proposal offers
important insights into the relation between the reduced phase space
quantization and the Dirac-Wheeler-DeWitt superspace formalism. The
author shows that at least for some choices of internal clocks, both
approaches are equivalent in a very well-defined sense. Specifically,
the author discusses in detail the relation between physical and
superspace propagators and inner products. However, the step of
constructing real observables in a gauge-independent way is left
out. It is not clear whether such a program can actually be achieved,
which is the reason for the time problem studied here; a simple and
general argument in favor of this position was given in
Ref.~\cite{hajicek1999choice}. Finally, another alternative approach
would involve arguing in favor of a preferred clock. We are not aware
of any widely recognized proposal of this type.

\begin{acknowledgements}
A.B. and P.M. acknowledge the support of the National Science Centre
(NCN, Poland) under the research Grant No.~2018/30/E/ST2/00370.
\end{acknowledgements}

\begin{figure}[b]
  \includegraphics[width=0.45\textwidth]{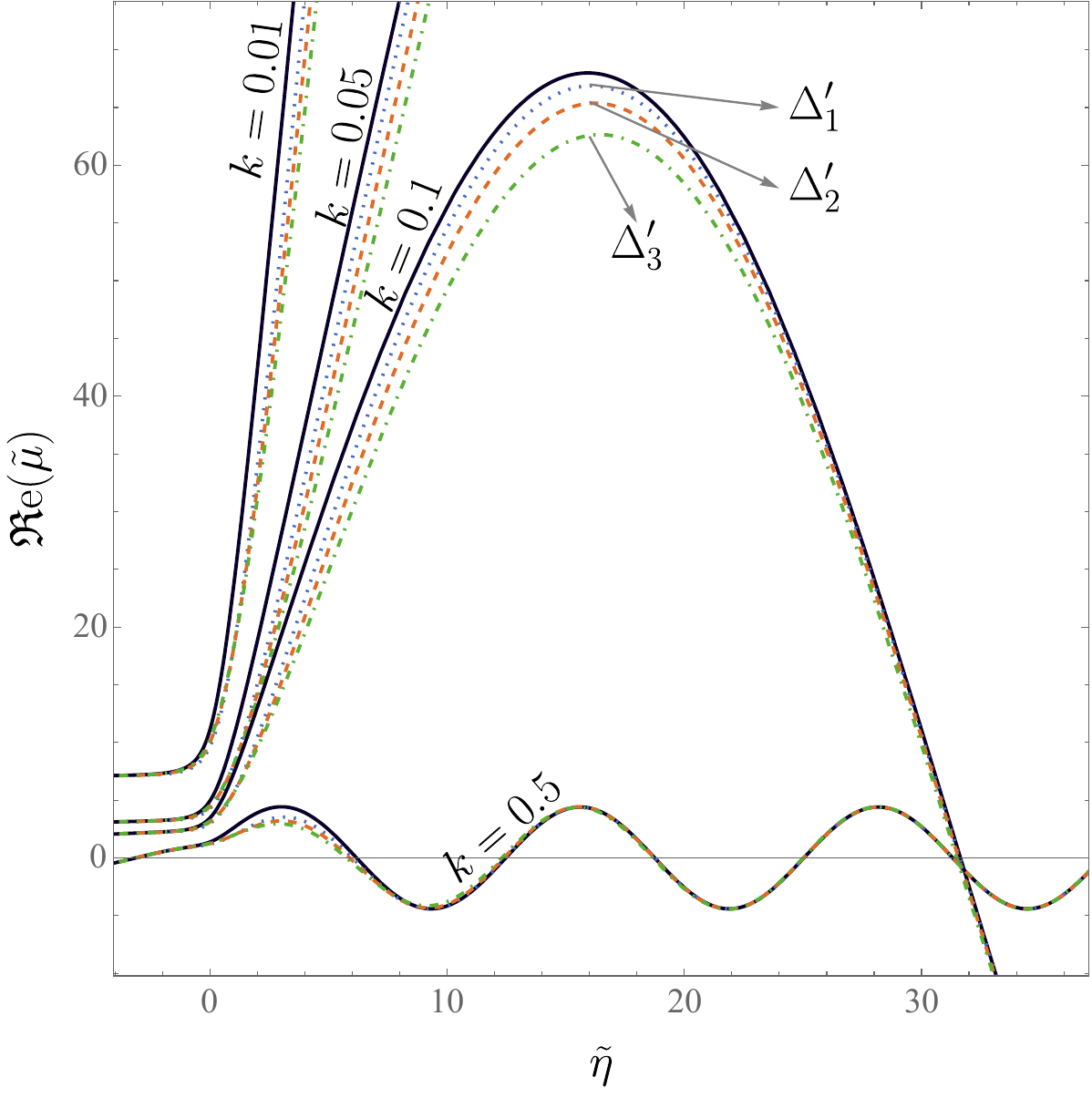}

  \caption{Evolution of the primordial gravitational amplitude for
    different clocks obtained from the second class of delay function
    \eqref{delfun2}. Convergence happens at a later time with respect
    to the first class of delay functions \eqref{delfun}, as can be
    seen by comparison with Fig.~\ref{mu-all-k}.}
  
  \label{mu-k-prime}
\end{figure}

\appendix*

\section{}\label{appro}

In this appendix, we consider the alternative choice of a family of
two-parameter delay functions, namely
\begin{equation}\label{delfun2}
\Delta'(a,p)=a^A e^{B p},
\end{equation}
which define a new set of clocks plotted in Fig.~\ref{deltasprime} for
a few relevant values of the parameters $A$ and $B$.
Figure~\ref{qpdeltaprime} depicts the trajectories with different
clocks obtained from $\Delta'(a,p)$ for which the convergence happens
much later than in the case discussed in the core of this paper, as
can be seen by comparing with Fig.~\ref{qpbackground}. The extent to
which this delay can be increased, and how the matter content of the
Universe can affect this limit, is not dealt with in the present
article and will be the subject of a future work.

One can note that the delay functions \eqref{deltasprime} tend to
diverge in time from one another, all of them growing exponentially
with the momentum; the phase space trajectories, however, do converge
to the undelayed one, but at scales that are increasingly larger with
the amplitude of the exponential behavior of the relevant delay
function.

Moving to the perturbations, we performed the same analysis as in the
core of this paper and show the time development of the real part of
the mode function for different values of the wave number in
Fig.~\ref{mu-k-clock2}, with a special emphasis at the near-bounce
regime in Fig.~\ref{mu-k-prime}. As for the other family of delay
functions, we find that whenever the classical approximation for the
background holds, one recovers a unique prediction.

\bibliography{references}

\clearpage
\newpage

\end{document}